\documentclass[12pt]{article}  
\usepackage[a4paper,margin=1in]{geometry}
\usepackage{dcolumn,color,helvet,times}
\usepackage{amsmath,lastpage,bm,textcomp}
\usepackage{graphicx}

\renewcommand{\eqref}[1]{Eq.~(\ref{#1})}

\newcommand{\PDF}[1]{p(#1)}
\newcommand{\CondPDF}[2]{p(#1 \mid #2)}
\newcommand{\like}[2]{\mathcal{L}(#1 ; #2)}

\newcommand{\V}[1]{\text{Var}\left[#1\right]}

\newcommand{\paramvec}{\boldsymbol{\theta}}

\newcommand{\dat}{\mathcal{D}}

\begin{document}


\title{Optimal quantification of contact inhibition in cell populations}

\author
{David J. Warne${}^{1}$, Ruth E. Baker${}^{2}$, Matthew J. Simpson${}^{1\ast}$\\
	\\
	\normalsize{${}^{1}$School of Mathematical Sciences, Queensland University of Technology,}\\
	\normalsize{Brisbane, Queensland 4001, Australia}\\
	\normalsize{${}^{2}$Mathematical Institute, University of Oxford,}\\
	\normalsize{Oxford, OX2 6GG, United Kingdom}
	\\
	\normalsize{$^\ast$To whom correspondence should be addressed; E-mail:  matthew.simpson@qut.edu.au.}
}

\date{\today}

\maketitle


\begin{abstract}
{Contact inhibition refers to a reduction in the rate of cell migration and/or cell proliferation in regions of high cell density. Under normal conditions contact inhibition is associated with the proper functioning tissues, whereas abnormal regulation of contact inhibition is associated with pathological conditions, such as tumor spreading.  Unfortunately, standard mathematical modeling practices mask the importance of parameters that control contact inhibition through scaling arguments.  Furthermore,  standard experimental protocols are insufficient to quantify the effects of contact inhibition because they focus on data describing early time, low-density dynamics only.  Here we use the logistic growth equation as a caricature model of contact inhibition to make recommendations as to how to best mitigate these issues.  Taking a Bayesian approach we quantify the trade-off between different features of experimental design and estimates of parameter uncertainty so that we can re-formulate a standard cell proliferation assay to provide estimates of both the low-density intrinsic growth rate, $\lambda$, and the carrying capacity density, $K$, which is a measure of contact inhibition.} 
\end{abstract}

Contact inhibition is the tendency of cells to become non-migratory and/or non-proliferative in regions of high cell density~\cite{Abercrombie1970}. The phenomena of contact inhibition of migration, involving processes such as adhesion, paralysis and contraction~\cite{Abercrombie1970}, is distinct to contact inhibition of proliferation, driven by cell-cell signaling and adhesion~\cite{Levine1964,Liu2011}; both phenomena are essential for the regulation of structure and function of multicellular organisms.

Down-regulation of contact inhibition of proliferation enhances tumor spreading~\cite{Abercrombie1979}, while wound healing and tissue regeneration also depend crucially on contact inhibition of proliferation~\cite{Puliafito2012}.  While contact inhibition of proliferation is ubiquitous in both normal and pathological processes, it is difficult to quantify the impact of such contact inhibition in complex biological systems despite the availability of experimental data. Therefore, mathematical models have an important role in informing our understanding of how contact inhibition of proliferation affects collective cell behavior.

The most fundamental mathematical model describing contact inhibition of cell proliferation is the logistic growth model~\cite{Maini2004b,Maini2004,Treloar2014,Tsoularis2002},
\begin{equation}
\dfrac{\textrm{d}{C}(t)}{\textrm{d}t} = \underbrace{\lambda C(t)}_{\textbf{proliferation}} \times \underbrace{\left[1-\dfrac{C(t)}{K}\right]}_{\textbf{contact inhibition}},
\label{eqn:lg}
\end{equation}
where $C(t)>0$ is the cell density at time $t$, $\lambda > 0$ is the proliferation rate, and  $K > 0$ is the carrying capacity density.  The carrying capacity density is the density at which contact inhibition decreases the net growth rate to zero.  The logistic growth model is used ubiquitously in the study of development, repair and tissue regeneration, including for the modeling of tumor growth~\cite{Enderling2014,Gerlee2013,Scott2013,Treloar2013} and wound healing~\cite{Maini2004b,Sherratt1990,Simpson2013}.
The solution of \eqref{eqn:lg},
\begin{equation}
C(t) = \dfrac{C(0)K}{\left[(K-C(0)) \exp(-\lambda t) + C(0)\right]},
\label{eqn:sol}
\end{equation}
is a sigmoid curve that increases from $C(t) = C(0)$ to $C(t) = K$ as $t \to \infty$, provided that $C(0)/K \ll 1$.

\emph{In vitro} cell proliferation assays are used routinely to examine mechanisms that control cell proliferation, such as the application of various drugs and other treatments on the rate of cell proliferation~\cite{Chen2017,Delarue2014}. \emph{In vitro} assays are routinely used to inform the development and interpretation of \emph{in vivo} assays describing pathological situations, such as tumor growth~\cite{Beaumont2014}.  Therefore, improving the design and interpretation of \textit{in vivo} assays will have an indirect influence on the way that we design and interpret \textit{in vivo} assays.

A cell proliferation assay typically involves placing cells, at low density, $C(0)/K \ll 1$, onto a two-dimensional surface and re-examining the increased monolayer density at a later time, $t = T$. Typical data from a cell proliferation assay are given in Fig. 1 A-B.

\begin{figure}
	\centering
	\includegraphics[width=0.55\columnwidth]{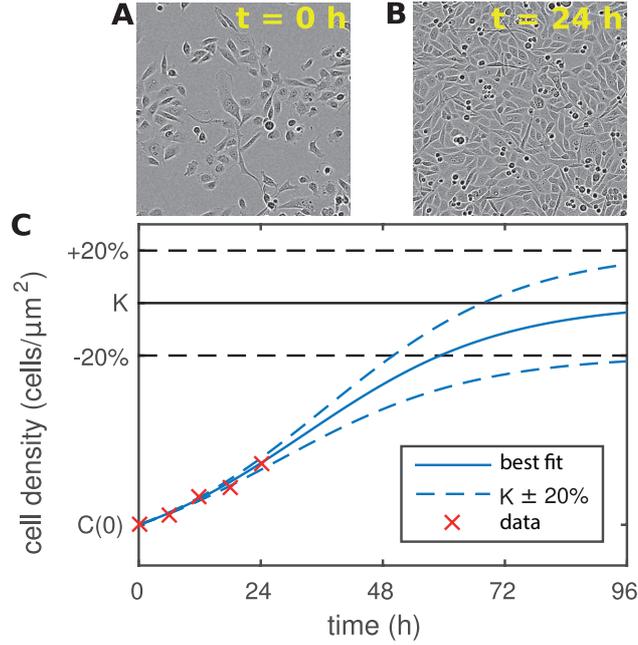}
	\caption{Proliferation assay using a prostate cancer cell line. Images of area 450 $\mu\text{m}^2$ are captured at (A) $t = 0 \,\text{ h}$ and (B) $t = 24\, \text{ h}$.  Images are reproduced from Jin et al., (2017), with permission. (C) Logistic growth curve with $C(0) = 3.1 \times 10^{-4}$ $(\text{cells}/\mu\text{m}^2)$, $\lambda = 0.052$ $(1/\text{h})$ and $K = 2 \times 10^{-3}$ $(\text{cells}/\mu\text{m}^2)$ (solid black).  Additional solutions with $K \pm 20\%$ (dashed black) also fit the short time experimental data (red crosses).}
	\label{fig:expr}
\end{figure}
To use \eqref{eqn:lg} to quantitatively inform our understanding of a particular biological system, we must be able to estimate the initial density, $C(0)$, and the model parameters, $\lambda$ and $K$.  Obtaining an accurate estimate of $K$ is crucial to understand how contact inhibition controls net the proliferation rate at modest to high densities. However, despite the importance of $K$, most theoretical studies work with a non-dimensionalized model by setting $c(t) = C(t)/K$.  This leads to~\cite{Maini2004b,Maini2004,Sherratt1990}
\begin{equation*}
\dfrac{\textrm{d} c(t)}{\textrm{d} t} = \lambda c(t) \left[ 1- c(t) \right],
\end{equation*}
which completely masks the importance of being able to accurately estimate the carrying capacity, $K$.

Not only do standard mathematical approaches prevent quantitative assessment of the impact of contact inhibition, in addition, standard experimental protocols for \textit{in vitro} proliferation assays are also insufficient to  estimate $K$.  Very recently, Jin et al.,~\cite{Jin2017} used \eqref{eqn:lg} to analyze a set of cell proliferation assays performed with a prostate cancer cell line, and concluded that standard experimental data do not lead to robust, biologically relevant estimates of $K$.   This is consistent with the work of Sarapata and de~Pillis who also find that standard \emph{in vitro} experimental data are insufficient to estimate $K$ using best-fit, non-linear least-squares methods~\cite{Sarapata2014}. The fundamental issue, as illustrated  in Fig.~\ref{fig:expr}C, is that cell proliferation assays are initiated with a small density and performed for a relatively short duration.  This strategy is sufficient for estimating the low-density intrinsic proliferation rate, $\lambda$, but completely inadequate for estimating $K$ which is associated with longer time, higher density data.  Given the importance of $K$, we are motivated to re-consider the design of proliferation assays so that we can quantitatively estimate both $\lambda$ and $K$ from a single experiment.

The typical duration of a proliferation assay is less than $24 \text{ h}$ \cite{Chen2017}, with some assays as short as $4 \text{ h}$ \cite{Huang2017}, and the cell density is recorded once, at the end point of the experiment.  This timescale is sufficient to estimate the low-density intrinsic proliferation rate, $\lambda$, since the doubling time of the majority of cell lines is approximately 24 hours~\cite{Maini2004b,Maini2004,Treloar2013,Simpson2013}.  However, this standard timescale is far too short to robustly quantify contact inhibition effects given the low initial densities that are routinely used. For example, the PC-3 prostate cancer cell line examined by Jin et al.~\cite{Jin2017} proliferates with $\lambda \approx 0.05$ $/$h, but on the timescale of the experiment the observed  growth of the population is approximately exponential.  This is consistent with \eqref{eqn:lg} since we have $C(t) \sim C(0)\exp(\lambda t)$ provided that $C(0)/K \ll 1$ and $t$ is sufficiently small. Fig.~\ref{fig:expr}C shows that the early time growth dynamics are effectively independent of $K$, confirming that it is impossible to obtain robust estimates of $K$ using standard data~\cite{Jin2017,Sarapata2014}.

The aim of this work is to provide guidance about how to overcome these standard experimental and theoretical limitations by taking a Bayesian approach to experimental design~\cite{Gelman2014,Liepe2013,Silk2014,Vanlier2012,Vanlier2014,Browning2017}.  The advantage of taking a Bayesian approach is that we have a platform to quantitatively examine how the uncertainty in our estimate of $K$ depends on the experimental design: we aim to provide guidance for experimental design that minimizes the uncertainty in our estimate of $K$.  To achieve this we consider a proliferation assay of duration $T$, with $n$ observations of cell density, $C_{\text{obs}}^{1:n} = [C_{\text{obs}}(t_1),\ldots, C_{\text{obs}}(t_n)$], at times $t_1,\ldots, t_n$ with $0 < t_i \leq T$ for all $i = [1,\ldots, n]$. These data could represent a single experiment observed at multiple time points, $t_1 < t_2 <\cdots < t_n \leq T$, or $n$ identically prepared experiments, each of which is observed once, $t_1 = t_2 = \cdots = t_n = T$. We assume that cells proliferate according to  \eqref{eqn:lg} with known $\lambda$ and $C(0)$.  Furthermore, we assume $C(0)/K \ll 1$. Our assumption that $\lambda$ can be determined is reasonable since $C(t) \sim C(0)\exp(\lambda t) = C(0)[ 1 + \lambda t + \mathcal{O}(t^2)]$ for $C(0)/K \ll 1$.  This means that fitting a simple straight line or exponential curve to typical experimental data will provide a reasonable estimate of $\lambda$. The assumption that $C(0)$ is known precisely is less realistic.  For example, estimates of $C(0)$ are affected profoundly by fluctuations in the initialization of the experiment as proliferation assays are performed by placing a known number of cells onto a tissue culture plate.  However, images of the experiments are obtained over a much smaller spatial scale.  This means that the role of stochastic fluctuations can be significant~\cite{Jin2017}.

To make progress we assume that observations are made, and are subject to Gaussian-distributed experimental measurement error with zero mean and variance $\Sigma^2$.  Under these conditions our knowledge of the carrying capacity, $K$, given such observations is represented by the probability density function
\begin{equation}
p\left(K\,\middle|\,C_{\text{obs}}^{1:n}\right)= A \prod_{i=1}^n \phi\left(C_{\text{obs}}(t_i); C(t_i),\Sigma^2\right),
\label{eqn:postmv}
\end{equation}
where $A$ is a normalization constant and $\phi\left( \cdot ;C(t_i),\Sigma^2\right)$ denotes a Gaussian probability density with mean $C(t_i)$ and variance $\Sigma^2$ (Supporting Material). This probability density represents knowledge obtained from the data when no prior assumptions are made on $K$. The point of maximum density in~\eqref{eqn:postmv} corresponds to the maximum likelihood estimator or best-fit estimate, $\hat{K}$. Importantly, \eqref{eqn:postmv} also enables the quantification of uncertainty in this estimate through calculation of the variance,
\begin{equation}
\sigma_n^2 = \int_{0}^{\infty} \left(K - \hat{K}\right)^2p\left(K\,\middle|\,C_{\text{obs}}^{1:n}\right)\,\text{d}K.
\end{equation}

\begin{figure}[t]
	\centering
	\includegraphics[width=0.55\columnwidth]{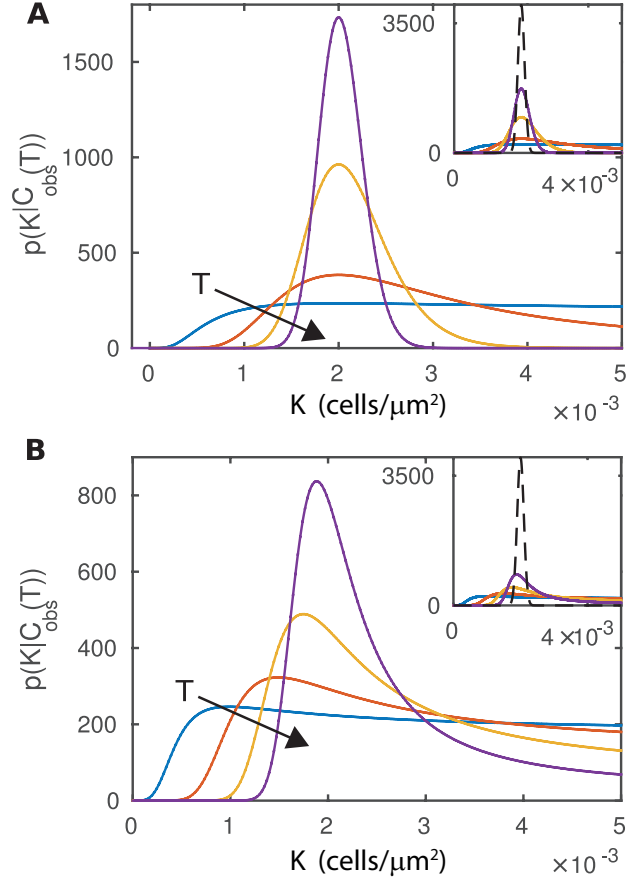}
	\caption{The probability density, $p\left(K\,\middle|\,C_{\text{obs}}^{1:n}\right)$, with $n = 1$ plotted for $T = 12$ h (black), $24$ h (red), $36$ h (yellow) and $48$ h (purple), where $\lambda = 5.2 \times 10^{-2}$ $(1/\text{h})$, $\sigma = 10^{-4}$ $(\text{cells}/\mu\text{m}^2)$ and the the true carrying capacity is $K = 2\times 10^{-3}$ $(\text{cells}/\mu\text{m}^2)$. (A) The initial density, $C(0)$, is assumed to be known precisely, $C(0) = 3.1 \times 10^{-4}$ $(\text{cells}/\mu\text{m}^2)$.  (B) Including uncertainty in the initial condition with $\mu_0 = 3.1 \times 10 ^{-4}$ $(\text{cells}/\mu\text{m}^2)$ and $\Sigma_0 = 1.02 \times 10 ^{-4}$ $(\text{cells}/\mu\text{m}^2)$. The inset panels in both (A) and (B) show the main plot in the context of the limiting density as $T \rightarrow \infty$ (black dashed line)}.
	\label{fig:conv}
\end{figure}

Fig.~\ref{fig:conv}A shows the probability density of $K$ (\eqref{eqn:postmv}) for several values of $T$ for the typical assay protocol where only a single observation is made ($n = 1$). Here, our estimates of $\lambda$, $C(0)$ and $\Sigma$ are based on reported values~\cite{Jin2017}. The spread of these curves indicates the degree of uncertainty in any estimate of $K$.  In particular, note the red line, which indicates the probability density for $K$ for a measurement taken at the standard duration of $T= 24 \text{ h}$. The relatively flat, disperse nature of the profile confirms that standard proliferation assay designs are completely inappropriate to estimate $K$ since the profile lacks a well-defined maximum.  This result provides a formal explanation for the observations of both Sarapata and de~Pillis~\cite{Sarapata2014} and Jin et al.~\cite{Jin2017}. In response to this issue, here we provide quantitative guidelines about how the experimental design can be chosen to facilitate accurate quantification of the effects of contact inhibition.

One optimistic assumption in \eqref{eqn:postmv} is the supposition that $C(0)$ is known precisely. In reality, $C(0)$ is subject to both measurement errors and systematic errors owing to stochastic fluctuations~\cite{Jin2017}. We extend our analysis to incorporate uncertainty in the estimate of $C(0)$ by also assuming $C(0)$ to be Gaussian-distributed with mean $\mu_0$ and variance $\Sigma_0^2$, that is $p(C(0)) = \phi\left(C(0);\mu_0,\Sigma_0^2\right)$. In this case \eqref{eqn:postmv} generalizes to (Supporting Material)
\begin{equation}
p\left(K\,\middle|\,C_{\text{obs}}^{1:n},\mu_0,\Sigma_0^2\right) = \int_0^{\infty}p\left(K\,\middle|\,C_{\text{obs}}^{1:n}\right) p(C(0))\, \text{d}C(0).
\label{eqn:postmvric}
\end{equation}
The integral in \eqref{eqn:postmvric} is intractable, so numerical integration is required to evaluate $p\left(K\,\middle|\,C_{\text{obs}}^{1:n},\mu_0,\Sigma_0^2\right)$.

Since $C(t_i) \to K$ as $t_i \to \infty$ for all $i = 1,2,\ldots,n$, then, for both \eqref{eqn:postmv} and \eqref{eqn:postmvric}, we obtain (Supporting Material)
\begin{equation}
\lim_{(t_1,\ldots,t_n) \to (\infty,\ldots,\infty)} p\left(K\,\middle|\,C_{\text{obs}}^{1:n}\right) = \phi\left(K; \frac{1}{n}\sum_{i=1}^n C_{\text{obs}}^i,\frac{\Sigma^2}{n}\right).
\label{eqn:postmvlim}
\end{equation}
Fig.~\ref{fig:conv}A demonstrates that the probability density function for $K$, \eqref{eqn:postmv}, tightens toward the limiting density, \eqref{eqn:postmvlim}, as $T$ increases. Note that for the typical assay duration, $T < 24 \text{ h}$, the density is approximately uniform.  Again, this reiterates the fact that standard experimental designs are inadequate to characterize the effects of contact inhibition, and that different approaches are required.  The effect of uncertainty in $C(0)$ is clear in Fig.~\ref{fig:conv}B. Even with $T = 48 \text{ h}$ there is a very large region of non-zero, near-constant probability density, indicating very wide confidence intervals for the estimate of $K$. \eqref{eqn:postmvlim} also provides a lower bound on the uncertainty in the estimate of $K$,
\begin{equation}
\sigma_n^2 \geq \dfrac{\Sigma^2}{n},
\label{eqn:uncertlb}
\end{equation}
for any choice of $C(0)$ and $\lambda$. This result is independent of the treatment of the uncertainty in $C(0)$, but requires observations to be made after an infinite duration of time.

\eqref{eqn:postmvlim} and \eqref{eqn:uncertlb} tell us two important things about assay design in the study of contact inhibition. First, there is a fundamental lower bound on the uncertainty in our estimate of $K$ for a fixed number, $n$, of observations of the density.  However, increasing the assay duration, $T$, always provides more information. Second, increasing the number of observations, $n$, always provides more information and, further, it decreases the long time lower bound on the uncertainty in the estimate of $K$. Hence, \eqref{eqn:uncertlb} informs the minimum number of observations required to estimate $K$ accurately.

\begin{figure}
	\centering
	\includegraphics[width=0.55\columnwidth]{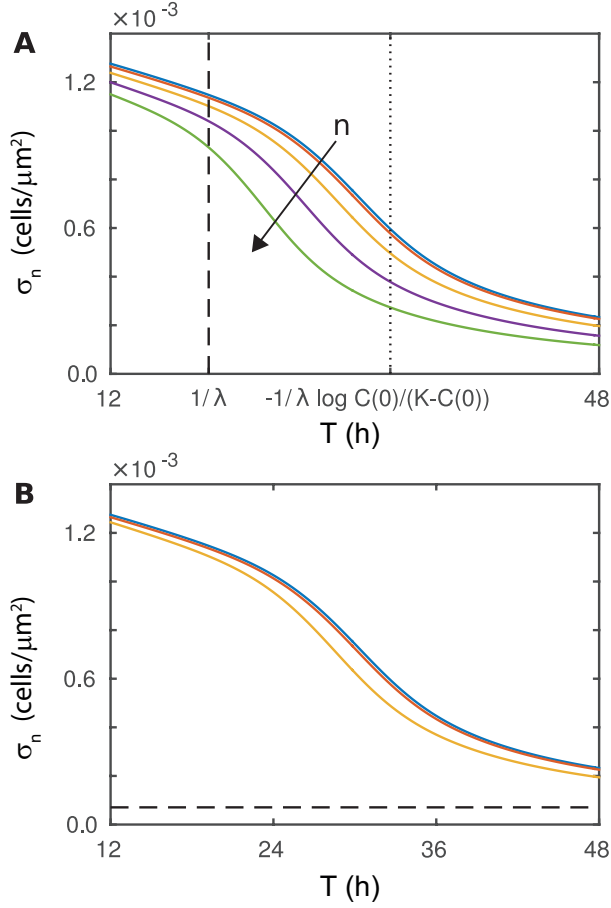}
	\caption{(A) Uncertainty in $K$ as a function of $T$ for $n = 1$ (black), $n = 2$ (red), $n = 4$ (yellow), $n = 8$ (purple) and $n = 16$ (green); here observations are taken at regular intervals. (B) Effect of observation placement for $n = 2$ with $t_2 = T$; $t_1 = T/5$ (black), $t_1 = T/2$ (red) and $t_1 = 4T/5$ (yellow). The lower bound on uncertainty, $\sigma_n$, for $n = 2$ (dashed).  The uncertainty, $\sigma_n$, is calculated using the trapezoid rule with $10^5$ equally spaced panels over the interval $ 0 <K<5\times10^{-3}$ $(\text{cells}/\mu\text{m}^2)$. }
	\label{fig:place}
\end{figure}

Clearly, practical experimental designs require finite $T$, and so we require methods to determine $T$ such that the uncertainty in $K$ is sufficiently close to the lower bound. We can quantify, for the standard choice of $n = 1$, the approximate uncertainty in our estimate of $K$ (Supporting Material)~\cite{Ang2007}, given by
\begin{equation}
\sigma_1^2 = \frac{C(0)^4(1-\exp(-\lambda T))^2}{\left[C(0) - C_{\text{obs}}(T)\exp(-\lambda T)\right]^{4}}\Sigma^2.
\label{eqn:relsig}
\end{equation}
This estimate of the uncertainty in $K$ is  accurate provided $C_\text{obs}(T)/\Sigma^2 \gg 1$. This expression provides a practical tool to assess the information content of data, but also enables one to estimate whether $T$ is large enough that the uncertainty in  $K$ is sufficiently close to the lower bound (Supporting Material).

\eqref{eqn:uncertlb} gives us a method of quantifying the information gained by introducing more observations in the idealized case that $T$ is sufficiently large.  For example,  doubling $n$ will half the uncertainty. However, since practical limitations mean that $T$ is finite, this result does not always hold. Fig.~\ref{fig:place}A shows that the effect of doubling $n$ varies significantly depending on the choice of $T$.  Here, the $n$ observations are taken at regular intervals.  For example, if $T < 1/\lambda$,  increasing $n$ has almost no effect on the uncertainty, $\sigma_n^2$.  There is a similar negligible effect for $T > -1/\lambda \log C(0)/(K - C(0))$, which is the time corresponding to the point of inflection of \eqref{eqn:sol}.

These results highlight several subtle, but immensely important considerations.  For example, at sufficiently short times increasing $n$ has very benefit.  Similarly, at sufficiently large times increasing $T$ has little benefit.  The most useful result is that for intermediate times there is more value in increasing $T$ than increasing $n$.  Furthermore, if we wish to go beyond the standard experimental design where a single measurement is made at the end of the experiment, $t = T$, we might want to quantify the benefit of making a second observation at an earlier time, $t < T$, during the same experiment.  In this scenario, results in Fig.~\ref{fig:place}B show that the choice of time at which the second measurement is made can be important. Comparing Fig.~\ref{fig:place}A with Fig.~\ref{fig:place}B shows that a poor choice of the time for the second observation might not lead to any change than taking a single observation at $T$. However, selecting two well-placed observation times can be as informative as making four equally spaced observations.

We conclude by providing several guidelines for the design of cell proliferation assays:
\begin{enumerate}
	\item Reducing the uncertainty in $C(0)$ is crucial, especially if $ T < 48 \text{ h}$;
	\item \eqref{eqn:relsig} should be used with short timescale data to estimate the smallest value of $T$ that will result in acceptable uncertainty in $K$;
	\item On a short timescale, increasing $T$ is more informative than increasing $n$. However, if increasing $T$ is infeasible, the optimal strategy is to repeat the experiment $n$ times and make $n$ observations at time $T$. In many situations, $n$ will need to be large to account for the short timescale. Fig.~\ref{fig:place}A shows that even a 16-fold increase in $n$ is still unacceptable for $T < 12$.
\end{enumerate}

In this work we highlight certain aspects of assay design that are often neglected and unreported.  However, these features are critical if we are to quantify the role of crowding and contact inhibition of proliferation in populations of cells. The guidelines we propose allow us to provide the best estimates of $\lambda$ and $K$ using a single experiment, whereas standard experimental designs allow us to confidently estimate $\lambda$ only. Our results confirm that standard \emph{in vitro} experimental designs are well suited for estimating $\lambda$, but poorly suited for estimating $K$. A different approach is required to overcome this limitation, and here we provide quantitative guidance about designing \textit{in vitro} proliferation assays that can be used to estimate both $\lambda$ and $K$.  The situation is even more complex for \textit{in vivo} assays in which there are more experimental constraints and unknowns.   However, improving the way that we design and interpret \textit{in vitro} assays is relevant because these simpler experiments are routinely used in tandem with \textit{in vivo} assays due to the fact that they are cheaper and faster to perform than working in live tissues.

\section*{Author Contributions}
DJW, REB and MJS designed the research. DJW performed the research. DJW, REB and MJS contributed analytic tools. DJW, REB and MJS analyzed the data. DJW, REB and MJS wrote the paper.

\section*{ACKNOWLEDGEMENTS}
MJS thanks the Australian Research Council (DP170100474). REB thanks the Royal Society for a Royal Society Wolfson Research Merit Award, and the Leverhulme Trust for a Leverhulme Research Fellowship. We thank Michael Stumpf and two anonymous referees for helpful comments.

\appendix
\section*{APPENDIX}
\setcounter{equation}{0}
\renewcommand{\theequation}{A.\arabic{equation}}
\subsection*{Posterior probability density functions}
In Bayesian statistics, knowledge of unobserved model parameters, $\paramvec$, given the results of some experiment that results in data, $\dat$, is represented through a probability density function (PDF)~\cite{Gelman2014}. This PDF, $\CondPDF{\paramvec}{\dat}$, is called the \emph{posterior} and can be interpreted as ``the probability density of $\paramvec$ given observation $\dat$''. The posterior is derived through Bayes' Theorem,
\begin{equation}
\CondPDF{\paramvec}{\dat} = \frac{\like{\paramvec}{\dat} \PDF{\paramvec}}{\PDF{\dat}},
\label{eqn:bayes}
\end{equation}
where the \emph{prior} PDF, $\PDF{\paramvec}$, represents knowledge before the experiment, the \emph{likelihood} function, $\like{\paramvec}{\dat}$, determines the probability density of the experimental results, $\dat$, for a given set of parameter values and the \emph{evidence}, $\PDF{\dat}$, is the likelihood taken across all parameter values. In effect, $\like{\paramvec}{\dat}$ encodes assumptions of the model, $\PDF{\paramvec}$ encodes the assumptions on the parameters and $\PDF{\dat}$ acts as a normalization constant.
Our task is to derive the posterior PDF for the carrying capacity density, $K$, given noisy observations of an assumed logistic growth curve.

First we derive the likelihood. We represent the process of taking a cell density measurement, $C_{\text{obs}}(t)$, at time, $t$, as a Gaussian random variable with mean around the true cell density, $C(t)$ (Eq.~(2)), and variance $\sigma^2$. Therefore, the probability density of a single observation is a Gaussian PDF,
\begin{align}
\phi(C_{\text{obs}}(t); C(t),\sigma^2) &= \frac{1}{\sigma \sqrt{2\pi}}e^{-(C_{\text{obs}}(t) - C(t))^2/(2\sigma^2)} \notag \\
&= \frac{1}{\sigma \sqrt{2\pi}}e^{-\left(C_{\text{obs}}(t) - \frac{C(0)K}{(K - C(0))e^{-\lambda t} + C(0)}\right)^2/(2\sigma^2)}.
\label{eqn:singleobs}
\end{align}
Assuming $C(0)$ and $\lambda$ are known, the exact logistic growth curve is fully determined for any proposed value of $K$. As a result, the observations made at time $t_1$ and $t_2$, $C_{\text{obs}}(t_1)$ and $C_{\text{obs}}(t_2)$, are independent given this value of $K$. Therefore the likelihood function for $n$ observations, $C_{\text{obs}}^{1:n} = [C_{\text{obs}}(t_1),\ldots,C_{\text{obs}}(t_n)]$, is given by
\begin{equation}
\like{K}{C_{\text{obs}}^{1:n}} = \prod_{i=1}^n \phi(C_{\text{obs}}(t_i); C(t_i),\sigma^2),
\label{eqn:like}
\end{equation}
where $\phi(C_{\text{obs}}(t_i); C(t_i),\sigma^2)$ is given by \eqref{eqn:singleobs}.

For the prior, we assume $0 \leq K \leq K_{\text{max}}$  with equal probability for some $K_{\text{max}} < \infty$, that is $K$ is uniformly distributed on the interval $(0,K_{\text{max}})$. The PDF is
\begin{equation}
\PDF{K} = \begin{cases}
\dfrac{1}{K_{\text{max}}}, & \text{if } K \in [0,K_{\text{max}}], \\
0, & \text{otherwise}.
\end{cases}
\label{eqn:prior}
\end{equation}
This uniform prior distribution imposes minimal assumptions, as we impose no prior knowledge about the value of $K$ outside of some upper bound that could be arbitrarily large. For uninformative priors, the modes of the posterior correspond to the maximum likelihood estimator (MLE).

The evidence acts as a normalizing constant to ensure the product of \eqref{eqn:like} and \eqref{eqn:prior} is a true PDF. Therefore, we have
\begin{align}
\PDF{C_{\text{obs}}^{1:n}} &= \int_{-\infty}^{\infty} \like{K}{C_{\text{obs}}^{1:n}} \PDF{K} \, \text{d}K \notag \\
&= \frac{1}{K_{\text{max}}}\int_{0}^{K_{\max}} \prod_{i=1}^n \phi(C_{\text{obs}}(t_i); C(t_i),\sigma^2) \, \text{d}K,
\label{eqn:evidence}
\end{align}
which converges since \eqref{eqn:singleobs} is bounded and continuous on the closed interval $K \in [0,K_{\text{max}}]$.

Finally, we arrive at the posterior through substitution of \eqref{eqn:like}, \eqref{eqn:prior}, and \eqref{eqn:evidence} into Bayes' Theorem, \eqref{eqn:bayes},
\begin{align}
\CondPDF{K}{C_{\text{obs}}^{1:n}} = \begin{cases}
A\prod_{i=1}^n \phi(C_{\text{obs}}(t_i); C(t_i),\sigma^2), & \text{if } K \in [0, K_{\text{max}}], \\
0, & \text{otherwise}.
\end{cases}
\label{eqn:postfic}
\end{align}
where 
\begin{equation}
\dfrac{1}{A} = \int_{0}^{K_{\max}} \prod_{i=1}^n \phi(C_{\text{obs}}(t_i); C(t_i),\sigma^2) \, \text{d}K, 
\end{equation}
which is Eq.~(3) from the main text.

As stated in the main text, \eqref{eqn:postfic} assumes $C(0)$ is fixed. To extend the model to capture uncertainty in $C(0)$ we define $C(0)$ to be a Gaussian random variable with mean $\mu_0$ and variance $\sigma_0^2$. The PDF is $\PDF{C(0)} = \phi(C(0); \mu_0,\sigma_0^2)$. In this case, the logistic growth curve, forming the means of the observations, depends on this random variable. Therefore, consider the joint distribution
\begin{align}
\CondPDF{K,C(0)}{C_{\text{obs}}^{1:n},\mu_0,\sigma_0^2} &= \CondPDF{K}{C(0),C_{\text{obs}}^{1:n}}\CondPDF{C(0)}{C_{\text{obs}}^{1:n},\mu_0,\sigma_0^2}\notag \\
&=  \CondPDF{K}{C(0),C_{\text{obs}}^{1:n}}\phi(C(0); \mu_0,\sigma_0^2),
\label{eqn:joint}
\end{align}
where $\CondPDF{K}{C(0),C_{\text{obs}}^{1:n}}$ is simply \eqref{eqn:postfic} with the dependence on $C(0)$ made explicit. The desired posterior is a marginal density of \eqref{eqn:joint}, that is,
\begin{equation}
\CondPDF{K}{C_{\text{obs}}^{1:n},\mu_0,\sigma_0^2} = \int_{-\infty}^{\infty} \CondPDF{K,C(0)}{C_{\text{obs}}^{1:n},\mu_0,\sigma_0^2}\, \text{d}C(0).
\label{eqn:postric}
\end{equation}
Substitution of \eqref{eqn:joint} into \eqref{eqn:postric} results in Eq.~(4) in the main text. If expanded, the posterior is
\begin{equation*}
\CondPDF{K}{C_{\text{obs}}^{1:n},\mu_0,\sigma_0^2} =
A\int_0^{K_{\text{max}}}\phi(C(0); \mu_0,\sigma_0^2)\prod_{i=1}^n \phi(C_{\text{obs}}(t_i); C(t_i),\sigma^2)\,\text{d}C(0),
\end{equation*}
for $K \in [0,K_{\text{max}}]$ and $\CondPDF{K}{C_{\text{obs}}^{1:n},\mu_0,\sigma_0^2} = 0 $ otherwise.

\subsection*{Limiting uncertainty}
We now consider the posteriors \eqref{eqn:postfic} and \eqref{eqn:postric} in the limit as the observation times become infinitely large, that is $t_i \to \infty$ for  $i \in [1,\ldots,n]$. First, note that since
$\displaystyle{\lim_{t_i \to \infty} C(t_i) = K}$, it is also true that,
\begin{equation}
\lim_{t_i \to \infty} \phi(C_{\text{obs}}(t_i); C(t_i),\sigma^2) = \phi(C_{\text{obs}}^i; K,\sigma^2).
\end{equation}
That is, all observations are independent, identically distributed Gaussian random variables with mean $K$ and variance $\sigma^2$ in the limit. Since $\phi(C_{\text{obs}}^i;K,\sigma^2) = \phi(K;C_{\text{obs}}^i,\sigma^2)$, then we extend the domain of $K$ to $K \in (-\infty,\infty)$ and obtain
\begin{equation}
\lim_{(t_1,\ldots,t_n) \to (\infty,\ldots,\infty)}\CondPDF{K}{C_{\text{obs}}^{1:n}} = \dfrac{\prod_{i=1}^n \phi(K; C_{\text{obs}}^i,\sigma^2)}{\int_{-\infty}^{\infty}\prod_{i=1}^n \phi(K; C_{\text{obs}}^i,\sigma^2)\,\text{d}K}.
\label{eqn:limpostfic}
\end{equation}
Note that for any two Gaussian PDFs $\phi(X;\mu_1,\sigma_1)$ and $\phi(X;\mu_2,\sigma_2)$ it can be shown that
\begin{equation}
\phi(X;\mu_1,\sigma_1)\phi(X;\mu_2,\sigma_2) \propto \phi\left(X;\frac{\mu_1\sigma_2^2 + \mu_2 \sigma_1^2}{\sigma_1^2 + \sigma_2^2},\frac{\sigma_1^2\sigma_2^2}{\sigma_1^2 + \sigma_2^2}\right).
\label{eqn:gprod}
\end{equation}
Through some tedious algebraic manipulations, we apply \eqref{eqn:gprod} to \eqref{eqn:limpostfic} to obtain
\begin{equation}
\lim_{(t_1,\ldots,t_n) \to (\infty,\ldots,\infty)}\CondPDF{K}{C_{\text{obs}}^{1:n}} = \phi\left(K; \frac{1}{n}\sum_{i=1}^n C_{\text{obs}}^i, \frac{\sigma^2}{n}\right).
\label{eqn:lim}
\end{equation}
Note that equality holds in \eqref{eqn:lim} because the constants of proportionality in \eqref{eqn:gprod} cancel out through \eqref{eqn:limpostfic}.

Similar steps are applied to \eqref{eqn:postric} to obtain
\begin{align}
\lim_{(t_1,\ldots,t_n) \to (\infty,\ldots,\infty)}\CondPDF{K}{C_{\text{obs}}^{1:n},\mu_0,\sigma_0^2} &= \int_{-\infty}^{\infty}  \phi\left(K; \frac{1}{n}\sum_{i=1}^n C_{\text{obs}}^i, \frac{\sigma^2}{n}\right)\PDF{C(0)}\, \text{d}C(0) \notag \\
&= \phi\left(K; \frac{1}{n}\sum_{i=1}^n C_{\text{obs}}^i, \frac{\sigma^2}{n}\right)\int_{-\infty}^{\infty}\PDF{C(0)}\,\text{d}C(0) \notag \\
&=  \phi\left(K; \frac{1}{n}\sum_{i=1}^n C_{\text{obs}}^i, \frac{\sigma^2}{n}\right) \times 1 \notag \\
&= \lim_{(t_1,\ldots,t_n) \to (\infty,\ldots,\infty)}\CondPDF{K}{C_{\text{obs}}^{1:n}}.
\label{eqn:eqlim}
\end{align}
This results in Eq.~(5) of the main text.

It is important to note that in the derivation of \eqref{eqn:lim} and \eqref{eqn:eqlim} we have extended the domain of $K$ to be the entire real number line. The prior in this case is actually improper, thus, for $K \in (-\infty,\infty)$, the posteriors \eqref{eqn:postfic} and \eqref{eqn:postric} are not guaranteed to be PDFs. However, in the limiting case when $t_i \to \infty$ for  $i \in [1,\ldots,n]$, the result is a PDF. Therefore, this extension is justified so long as $K/\sigma \gg 1$. If this is not satisfied, then then this analysis could be completed using a truncated Gaussian prior.

\subsection*{Short time uncertainty quantification}
Here we derive Eq.~(7) of the main text. The approximation uses the delta method to give an approximate variance for \eqref{eqn:postfic}, that is the uncertainty in $K$ for $T < \infty$ and $n = 1$.

The posterior for $n = 1$ is
\begin{equation}
\CondPDF{K}{C_{\text{obs}}(T)} = \frac{1}{\sigma \sqrt{2\pi}}e^{-\left(C_{\text{obs}}(T) - C(T) \right)^2/(2\sigma^2)}.
\end{equation}
Thus, we can treat $C(T)$ as a Gaussian random variable with mean $C_{\text{obs}}(T)$ and variance $\sigma^2$. Given a realization of $C(T)$ we can obtain from the logistic growth solution (Eq. (2) in main text)
\begin{equation}
K = f(C(T)) = \frac{C(T)C(0)(1-e^{-\lambda T})}{C(0) - C(T)e^{-\lambda T}}.
\label{eqn:kfunc}
\end{equation}

The delta method is an approximation technique for obtaining approximate moments of a distribution of a random variable that is a function of another~\cite{Ang2007}. The idea is to consider the Taylor expansion about the mean of $C(T)$,
\begin{equation*}
f(C(T)) = f(C_{\text{obs}}(T) + \delta) = \sum_{k=0}^\infty \frac{\delta^k}{k!}\frac{\text{d}^k f}{\text{d}C^k}[C_{\text{obs}}(T)].
\end{equation*}
From this expansion we obtain
\begin{equation*}
\V{K} = \V{f(C(T))} = \V{\sum_{k=0}^\infty \frac{\delta^k}{k!}\frac{\text{d}^k f}{\text{d}C^k}[C_{\text{obs}}(T)]}.
\end{equation*}
For small $\delta$ we have
\begin{equation*}
\V{K} = \V{f(C_{\text{obs}}(T)) + \delta\frac{\text{d}f}{\text{d}C}[C_{\text{obs}}(T)] +\mathcal{O}(\delta^2)}.
\end{equation*}
Recall that $\V{a} = 0$, $\V{X + a} = \V{X}$ and $\V{aX} = a^2\V{X}$ for any random variable, $X$, and constant $a$. Therefore,
\begin{align}
\V{f(C_{\text{obs}}(T)) + \delta\frac{\text{d}f}{\text{d}C}[C_{\text{obs}}(T)] +\mathcal{O}(\delta^2)} &= \V{\delta\frac{\text{d}f}{\text{d}C}[C_{\text{obs}}(T)] +\mathcal{O}(\delta^2)} \notag\\
&= \left(\frac{\text{d}f}{\text{d}C}[C_{\text{obs}}(T)]\right)^2\V{C(T)} + \mathcal{O}(\mu_4),
\label{eqn:apprx1}
\end{align}
where $\mu_4$ is the fourth central moment of the distribution of $C(T)$. The approximation is appropriate provided $|\delta|$ is sufficiently small, that is $|C(T) - C_{\text{obs}}(T)|$ is sufficiently small.  This is achieved if $C_{\text{obs}}(T)/\sigma \gg 1$. The derivative of $f$ with respect to $C$ is
\begin{equation}
\frac{\text{d}f}{\text{d}C} = \frac{C(0)^2(1-e^{-\lambda T})}{[C(0) - C(T)e^{-\lambda T}]^2}.
\label{eqn:dfdc}
\end{equation}
Therefore, by substitution of \eqref{eqn:dfdc} into \eqref{eqn:apprx1},  we arrive at the approximation
\begin{align}
\V{K} \approx \frac{C(0)^4(1-e^{-\lambda T})^2}{[C(0) - C_{\text{obs}}(T)e^{-\lambda T}]^4}\sigma^2.
\label{eqn:apprx2}
\end{align}

There are two practical uses for \eqref{eqn:apprx2}. First, given an population density observation, $C_{\text{obs}}(T)$, taken at time $T$, one can calculate
\begin{equation*}
\epsilon = \frac{C(0)^4(1-e^{-\lambda T})^2}{[C(0) - C_{\text{obs}}(T)e^{-\lambda T}]^4}.
\end{equation*}
Here, $\epsilon$ provides a relative uncertainty compared to the limiting best case $\sigma$. Therefore, the test provides a measure of how close to optimal the assay is. Furthermore, if the observation error is known then this sample approach provides the uncertainty in $K$ after the observation. Second, given the upper limit of the prior $K_{\text{max}}$, as in \eqref{eqn:prior}, one may consider the function
\begin{equation*}
h(T) = \frac{C(0)^4(1-e^{-\lambda T})^2}{[C(0) - K_{\text{max}}e^{-\lambda T}]^4},
\end{equation*}
to identify how large $T$ must be to be close enough to the limiting posterior, i.e., $\epsilon \approx 1$. If this value of $T$ is too large for practical purposes, then it indicates more observations should be taken. It is important to note that the approximation does require $C_{\text{obs}}(T)/\sigma \gg 1$. In practice, this may not hold for early time, $T \ll 1/\lambda$. In such a case, \eqref{eqn:apprx2} tends to be an \emph{underestimate}. Because of this, the result is still useful to decide if increasing $n$ is valuable or not.

\subsection*{Proliferation data}

Table~\ref{tab:data} presents the data used to inform Fig. 1, Fig. 2 and Fig. 3 in the main text. The data are derived from Jin et al.~\cite{Jin2017}.

\begin{table}[h!]
	\caption{Cell density data}
	\begin{tabular}{l|ccccc}
		time (h) & 0 & 6 & 12 & 18  & 24 \\
		\hline
		cell density (cells/$\mu$m$^2$) & $3.1\times10^{-4}$ & $3.8\times10^{-4}$ & $5.2\times10^{-4}$ & $5.9\times10^{-4}$ & $7.8\times10^{-4}$
	\end{tabular}
	\label{tab:data}
\end{table}

\end{document}